\title{\c{C}inlar Subgrid Scale  Model for Large Eddy Simulation}

\documentclass[preprint,12pt]{elsarticle}
\usepackage{amssymb}
\usepackage{graphics}
\usepackage{graphicx}
\usepackage{epsfig}
\usepackage{amsmath}
\usepackage{dsfont}
\usepackage{booktabs}
\usepackage{amsfonts}
\usepackage{caption}
\usepackage{subcaption}

\begin{document}

\begin{frontmatter}



\title{\c{C}inlar Subgrid Scale Model for Large Eddy Simulation}




\author[rvt]{R.~Kara}
\ead{rukiye.kara@msgsu.edu.tr}
\author[focal]{M.~\c{C}a\u{g}lar}
\ead{mcaglar@ku.edu.tr}

\address[rvt]{Mimar Sinan Fine Arts University,Istanbul, Turkey}
\address[focal]{Koc University, Istanbul, Turkey}



\begin{abstract}

We construct a new subgrid scale (SGS) stress model  for representing the small scale effects in large eddy simulation  (LES) of incompressible flows. We use the covariance tensor for representing the Reynolds stress and include Clark's model for the cross stress. The Reynolds stress is obtained analytically from \c{C}inlar random velocity field, which is based on vortex structures observed in the ocean at the subgrid scale. The validity of the model is tested with turbulent channel flow computed in OpenFOAM. It is compared with the  most frequently used  Smagorinsky and one-equation eddy SGS models through DNS data.

\end{abstract}


\end{frontmatter}
\section{Introduction}

In fluid dynamics, the turbulent motion has chaotic and stochastic behaviour, which is modelled by the Navier-Stokes equations (NSE). On the other hand, the exact numerical or analytical solution of these equations is still open in turbulence theory. To solve NSE numerically, direct numerical simulation (DNS) is the most precise technique, which requires to include all the scales, small and large. Obviously, this procedure has a heavy  computational cost. The large eddy simulation (LES), which is based on modelling the effect of the small scales on the larger scales, is an efficient numerical solution method for NSE. In LES, by applying a filter,  the  large
scales (low frequency) are separated from the small scales (high frequency). After the filtering procedure,  a nonlinear term  called the subgrid scale stress (SGS) appears in LES equations.  Therefore,  the SGS term which remains unresolved should be  modelled. It is formed by  the  Reynolds stress, the cross stress and the Leonard stress. The  Reynolds stress consists of the fluctuations of the velocity representing the subgrid scales, the cross stress includes nonlinear interactions of the resolved and the subgrid scales, and the Leonard stress involve only the resolved velocity,  which is computed numerically.

To express a physically valid closure model for the SGS stresses is the basic difficulty in LES, and many models have been proposed.  Lilly \cite{lilly62, lilly67}, Deardorff
\cite{deardorff70}, Leonard \cite{leonard74} and Smagorinsky \cite{sma63} are among  the pioneers of the SGS models.  Most commonly used subgrid stress model is Smagorinsky, an eddy viscosity SGS model proposed by Smagorinsky \cite{sma63}. Moreover, assuming that  the smallest resolved
scales is similar to  the largest unresolved scales structurally,  Bardina et al.  \cite{bardina80} developed scale similarity model. Later, a dynamic eddy viscosity model, in which the eddy viscosity coefficient is computed dynamically, is introduced by Germano, Piomelli, Moin and Cabot \cite{germano91}.  See also \cite{sagaut} for an account of various related models. Different from the above models, Misra and Pullin \cite{misra97} developed a subgrid model based on stretched vortices where the orientation  of the vortices is determined by the resolved scales
and randomized parameters. They have taken the Reynolds stress as proportional to the energy spectrum that is expressed in terms of  the vortices.

Our aim  is to derive a new subgrid stress model using \c{C}inlar random velocity field, which is also based on vortex structures \cite{cinlar1993}. Its theory has been studied extensively
as a model for small to medium scale turbulent flow, see e.g. \cite{caglar2000,caglar2003,caglar2006,caglar2007}.
We have recently shown that the velocity field can capture the second order properties of the subgrid scale with its energy spectrum in \cite{kara2015} as a plausible turbulence model. \c{C}inlar velocity spectrum which is based on the truncated Gamma distribution indicates a good match with the spectrum estimated from real data, and it is similar to the widely used form of energy spectrum for small scales. Our initial attempts for modelling  Reynolds stress  have appeared in \cite{kara2014} where we proposed to model the energy spectrum like Misra and Pullin \cite{misra97}, but relying on \c{C}inlar  random velocity field as a physically valid model for the subgrid scale velocity involving vortex structures.

In this paper, we refine and develop the ideas of \cite{kara2014} for modelling Reynolds stress directly from the covariance tensor of \c{C}inlar velocity field, which is available in closed form, rather than the energy spectrum which would require more computations. Indeed, we have the analytical expression
\[
\frac{\lambda
}{2c}\mathbb{E}({{a}^{2}})\mathbb{E}({{b}^{2}})\int_{{{\mathbb{R}}^{2}}}dz\,{\frac{{{\upsilon
}_{i}}(z){{\upsilon }_{j}}(z)}{|z{{|}^{2\gamma }}}}
\]
for the covariance tensor, where and $\gamma>0$ is a constant, $\lambda$, $c$ are  parameters denoting the arrival rate per unit time-unit space and the decay rate of an eddy, respectively,  $a$ is the random amplitude and $b$ is the random radius  of an eddy, and $\upsilon$ is a  standardized eddy over $\mathbb{R}^2$ that other eddies are obtained by randomization. The covariance at space and time lag (0,0) is used as an approximation for the Reynolds stress. Then, the parameters $\lambda$, $c$, and the parameters originating from  the probability distributions of $a$ and $b$ are modelled as functions of the resolved strain rate  to approximate the Reynolds stress part of the SGS stress tensor. As a result, we obtain by physical and dimensional considerations an approximation of the Reynolds stress as
$$
R_{ij}\equiv \delta
_{ij}\frac{3\pi}{64}\frac{C_3^2C_4}{C_1}\bar{u}^2\bar{\Delta}^2e^{-2C_1|\bar{S}|}
 f(C_2\bar{\Delta}/(|\bar{S}|\bar{u}),C_3 )
$$
where $C_1, \ldots ,C_4$ are positive constants, $f$ is an explicit function, $|\bar{S}|$ is the magnitude of the resolved strain rate, $\bar{u}$ is the resolved velocity in LES obtained after filtering the Navier-Stokes equations, and $\bar{\Delta}$ is the grid size.

By extensive numerical computations, we compare  \c{C}inlar SGS model based on its particular Reynolds stress with two widely used SGS models, namely, Smagorinsky and one equation for Reynolds numbers $395, 590$ and $950$.  These benchmark models are available in OpenFOAM, which is    open source software for computational fluid dynamics \cite{openfoam}. We perform LES of
fully  developed  incompressible turbulent  channel  flow in OpenFOAM.
Although related, our model cannot be considered as an eddy viscosity model where only the viscosity parameter would be modelled. In our case, there is a more physical velocity model based on vortices, and the resultant model includes the parameters which are modelled with the strain rate.
As a result of the computational comparison, our model is shown to provide better approximation of the fluctuations in the viscous range than the benchmark models. Besides, it is numerically efficient with less computational cost.

The paper is organized as follows. In Section 2, a brief introduction is given about LES and the benchmark SGS models. In Section 3, \c{C}inlar velocity field is reviewed. In Section 4,  Reynolds stress is modelled using \c{C}inlar velocity field. Numerical results for turbulent channel flow are given in Section 5. Finally, Section 6 concludes the paper.


\section{LES and SGS Models}

\subsection{Large Eddy Simulation}
Large eddy simulation is a  numerical simulation technique for turbulent flows, where the effect of small scales is modelled.  LES is based on  decomposing flow variables into the resolved (filtered) and the unresolved subgrid scale terms. The velocity field can be decomposed as
$$u(x)=\bar{u}(x)+u'(x)$$
where
$$ \bar{u}(x)=\int_{-\infty}^{\infty}u(\xi)G(x-\xi)d^3\xi$$
is the filtered velocity field in space and  $G$ is the filter function that determines the size and structure of the small scales.

If the filtering operation is applied to Navier-Stokes equations for incompressible flows, the filtered Navier-Stokes equations are obtained as
\begin{eqnarray}
\frac{\partial \bar{u}_i}{\partial x_i}&=&0 \nonumber \\
\frac{\partial \bar{u}_i}{\partial t}+\frac{\partial (\bar{u}_i\bar{u}_j)}{\partial x_j}&=&-\frac{\partial \bar{p}}{\partial x_i}+\nu\frac{\partial}{\partial
x_j}\left(\frac{\partial \bar{u}_i}{\partial x_j}+\frac{\partial \bar{u}_j}{\partial x_i}\right)-\frac{\partial \tau_{ij}}{\partial x_j} \label{Les}
\end{eqnarray}
where $\tau_{ij}=\overline{u_iu_j}-\bar{u}_i\bar{u}_j$ is subgrid scale tensor and it must be modelled to represent the effect  of small scales.

Leonard \cite{leonard74} decomposed subgrid stress tensor as
$$ \tau_{ij}=\overline{u_iu_j}-\bar{u}_i\bar{u}_j=L_{ij}+C_{ij}+R_{ij}$$
and provided physical interpretations for each term. $L_{ij}=\overline{\bar{u}_i\bar{u}_j}-\bar{u}_i\bar{u}_j$, the so-called Leonard tensor, represents interactions among large scales and can be computed explicitly. $R_{ij}=\overline{u'_iu'_j}$, the Reynolds stress term, represents interactions among the small scales, and $C_{ij}=\overline{\bar{u}_iu'_j}+\overline{u'_j\bar{u}_i}$, the cross term, represents cross-scale interactions between the resolved and unresolved scales. Modelling the non-linear term $\tau_{ij}$ is the aim of subgrid scale (SGS) models.

\subsection{Subgrid Scale Models}
The SGS turbulence models usually use the eddy viscosity idea focusing on energy dissipation at subgrid scale based on Boussinesq's theory, which states that  the subgrid stress tensor is proportional to the resolved strain rate. Therefore, the deviatoric part of SGS stress tensor
is modelled as
\begin{equation}\nonumber
\tau_{ij}^d:=\tau_{ij}-\frac{1}{3}\tau_{kk}\delta_{ij}=-2\nu_{t}\bar{S}_{ij}
\end{equation}
 where
 \begin{equation}\label{strainrate}
 \bar{S}_{ij}=\frac{1}{2}\left(\frac{\partial \bar{u}_i}{\partial x_j}+\frac{\partial \bar{u}_j}{\partial x_i}\right)
 \end{equation}
is the resolved strain rate tensor, and $\nu_{t}$ is the eddy viscosity. In LES, the term $\frac{1}{3}\tau_{kk}\delta_{ij}$ is embedded in the pressure term  as $P=\bar{p}+\frac{1}{3}\tau_{kk}\delta_{ij}$ in \eqref{Les}, and only the deviatoric part $\tau_{ij}^d$ of the SGS stress tensor $\tau_{ij}$ is modelled.

Our SGS model, which is based on \c{C}inlar velocity field for subgrid scale, also exploits the approximation of viscous effects with the resolved strain tensor. Therefore, Smagorinsky and  one equation models are used as benchmark for comparison in the present work. Both are conveniently available in OpenFOAM, which is the open source software used in our computations, and are classified under eddy viscosity models.

The most common SGS model is  Smagorinsky model \cite{sma63}.  The eddy viscosity  is modelled by using the magnitude of strain rate tensor and the characteristic length scale. Characteristic length scale can be taken as proportional to the filter width with  Smagorinsky constant denoted by $C_S$.  Consequently, the eddy viscosity is given by
\begin{equation}
\nu_{t}=(C_s\bar{\Delta})^2|\bar{S}|
\end{equation}
where   $$|\bar{S}|^2=2\bar{S}_{ij}\bar{S}_{ij}$$  is the magnitude of the  strain  rate tensor, $\bar{\Delta}$ is the grid size.

Another popular eddy-viscosity model of similar form to  Smagorinsky closure relates $\nu_t$ to the subgrid scale turbulence kinetic energy  of the flow $k^{sgs}$ as \cite{deardoff1980}
 $$\nu_t=C_v\bar{\Delta}\sqrt{k^{sgs}}$$ 
where $C_v$ is a model constant. The  turbulence kinetic energy based  approach known as one equation model requires solving an extra equation for the subgrid scale kinetic energy. The transport equation for $k^{sgs}$ is given by \cite{schumann, horiuti, kim}:
\begin{equation}\nonumber
\frac{\partial k^{sgs}}{\partial t}+\frac{\partial}{\partial x_i}(\bar{u}_ik^{sgs})=-\tau_{ij}\frac{\partial \bar{u}_i}{\partial x_j}-C_v\frac{(k^{sgs})^{3/2}}{\bar{\Delta}}+\frac{\partial}{\partial x_i}\left(\nu_t\frac{\partial k^{sgs}}{\partial x_i}\right).
\end{equation}

\section{Subgrid Velocity Field}

We consider \c{C}inlar velocity field, which has  been motivated by subgrid scale observations and shown to represent its statistical properties very well \cite{caglar2006,caglar2007}. Let $\upsilon$ be a deterministic velocity field on $\mathds{R}^2$
called the basic eddy, and let $Q =
\mathds{R}^2\times\mathds{R}\times(0,\infty)$ be  the set of types of eddies.
Eddies of different sizes and amplitudes for $q \in Q$ , $x
\in\mathds{R}^2$ are obtained by
   \begin{equation}\nonumber
      \upsilon_q(x) = a\, \upsilon \left(\frac{x-z}{b}\right), \quad q=(z,a,b)
   \end{equation}
where $q$ represents the type of an eddy and includes its center $z$
in space, its amplitude $a$ as well as its radius $b$. Let $N$ be a
Poisson random measure on the Borel sets of $\mathds{R}\times Q$
with mean measure
\begin{equation}\nonumber
     \mu(dt, dq) \equiv \mu(dt,dz,da,db) =
     \lambda \, dtdz\alpha(da)\beta(db)
\end{equation}
where $\lambda$  is the arrival rate per unit time-unit space, and
$\alpha$ and $\beta$ are probability distributions for the amplitudes and radii of eddies, respectively. The
arrival time $t$ of an eddy, its center $z$, amplitude $a$ and
radius $b$ are all randomized with $N$. By the superposition of these
eddies decaying exponentially in time with rate $c_q$, which depends on the type $q$ of an eddy,  the generalized form of \c{C}inlar velocity field is constructed as
     \begin{equation}\label{velocity}
       u'(x,t)=\int_{-\infty}^t\int_Q
       N\left(ds,dz,da,db\right)e^{-c_q (t-s)} a\, \upsilon\left(\frac{x-z}{b}\right)
    \end{equation}
where $x\in \mathds{R}^2$, $t\in\mathds{R}$, and the notation $u'$ is used to indicate that we aim to model the subgrid scales with \eqref{velocity}.
The decay parameter is explicitly given by
    \begin{equation}\nonumber
        c_q(x)=c\left|\frac{x-z}{b}\right|^{2\gamma}
    \end{equation}
for $q=(z,a,b)$, where $c>0$ and $\gamma > 0$ \cite{caglar2007}.

The construction of \c{C}inlar velocity field is motivated from
vortex development and decay observed in the ocean \cite{shay2000}.
Therefore, we consider an incompressible and isotropic flow   in $\mathds{R}^2$ by taking the basic eddy $\upsilon = (\upsilon_1, \upsilon_2)$ as a
rotation around 0 with magnitude $m(r)$ at distance $r$ from 0, where $m :\mathds{R}\rightarrow\mathds{R}_+$ is continuous and has support $[0, 1]$. In particular,  $m(r)=(1-\cos 2\pi r)/{2},\quad 0\leq  r\leq 1$, and $m(r)=0$ otherwise. The specific expressions for $\upsilon$ are
    \begin{equation}\label{basicvelocity}
       \upsilon_1\left(x\right)=-\frac{x_2}{r}m(r), \quad \upsilon_2\left(x\right)=\frac{x_1}{r}m(r)
    \end{equation}
 where $x = (x_1, x_2)$ and $r = |x|\in [0, 1]$.

The covariance tensor of the velocity field can be computed analytically as
     \begin{eqnarray}\label{c1}
       R^{ij}\left(x,t\right)=\frac{\lambda}{c}\int_{\mathds{R}^2}
       dz\int_{\mathds{R}}\alpha(da)a^2\int_{\mathds{R}+}&& \!\!\!\!\!\!\!\!\!\!\!\!
       \beta(db)\frac{b^2\exp{\left(-c|z|^{2\gamma}|t|\right)}}{|z|^{2\gamma}+|z+\frac{x}{b}|^{2\gamma}}\\ \nonumber
       &\cdot&\upsilon_i\left(z\right)\upsilon_j\left(z+\frac{x}{b}\right)
     \end{eqnarray}
for $x\in \mathds{R}^2$ and $t\in \mathds{R}$, where the time integral has already been taken. We will consider only small scale eddies
up to some cutoff $B$. Therefore, the distribution $\beta$ of $b$ is chosen as a
right-truncated Gamma distribution given by
    \begin{equation}\label{gamma}
    \beta\left(db\right)=\frac{b^{\theta-1}\exp{\left(-b/\zeta\right)}}
    {\Gamma_{B/\zeta}\left(\theta\right)\zeta^\theta}\,db, \quad 0<b<B
    \end{equation}
where $\theta>0$ and $\zeta >0$ are the shape and scale parameters,
respectively, and $\Gamma_{B/\zeta}\left(\theta\right)$ is the
incomplete Gamma function with parameter $\theta$ and integration
bounds from 0 to $B/\zeta$. The energy spectrum has been obtained from the Fourier transform of $R$ with truncated Gamma distribution and studied in \cite{kara2015} for further validating \c{C}inlar velocity as a plausible turbulence model.

\section{Modelling Reynolds Stress}
Modelling the subgrid stress tensor $\tau_{ij}$ is the key step of  LES. As a term in the filtered Navier-Stokes equation, $\tau_{ij}$ reflects the effect of small scales on large scales. Recall that it is decomposed as
\begin{equation}\nonumber
\tau_{ij}=L_{ij}+C_{ij}+R_{ij}.
\end{equation}
In this section, we explain how the Reynolds stress $R_{ij}$ is obtained and modelled. For the cross stress $C_{ij}$, we use Clark's cross stress model. Clark \cite{clark1977} has modelled cross stress using Taylor series expansion as
\begin{equation}\label{cross}
\overline{\bar{u}_iu'_j}=\frac{\bar{\Delta}^2}{24}\bar{u}_i\frac{\partial^2\bar{u}_j}{\partial x^2_k}+O(\bar{\Delta}^4)
\end{equation}
in terms of the resolved scales. On the other hand, Leonard stress $L_{ij}$ does not need to be modelled as it depends only on the resolved velocity field.

\subsection{Reynolds Stress from Subgrid Velocity}
Homogeneity and isotropy properties of turbulence indicate that the statistical properties of fluctuations $ u'$ are  independent of the position and orientation. In addition, if the statistical  properties do not depend on time, the random field is called stationary. So, the covariance tensor in space and time is given by
\[
R^{ij}(x,t):= \mathbb{E}[u'_i(r,s)u'_j(r+x,s+t)]
\]
for two-point velocity. Clearly,  $R^{ij}(x,t)$ does not depend on the point $r$ in space and the time $s$ for homogeneous and stationary turbulence.
 The covariance function of \c{C}inlar velocity field is computed as
\begin{eqnarray*}
\lefteqn{{{R}^{ij}}(x,t)=\frac{\lambda }{c}\int_{\mathbb{R}}{\alpha
}(da){{a}^{2}}\int_{{{\mathbb{R}}^{2}}}{d}z\exp (-c|z{{|}^{2\gamma
}}|t|)}\\
&\quad \quad \quad \quad \displaystyle{\cdot \int_{{{\mathbb{R}}^{+}}}{}db\frac{{{b}^{\theta -1}}\exp
(-b/\eta )}{\Gamma (\theta ){{\eta }^{\theta }}}\frac{{{\upsilon
}_{i}}(z){{\upsilon }_{j}}(z+\frac{x}{b})}{|z{{|}^{2\gamma
}}+|z+\frac{x}{b}{{|}^{2\gamma }}}}
\end{eqnarray*}

 Reynolds stress represents the interaction of  small scales. In Reynolds averaged Navier-Stokes equation, Reynolds stress is defined as a time average. Because  time averages converge to statistical averages by stationarity, and ergodicity when applicable, Reynolds  stress is modelled as the covariance of the subgrid velocity field.

A subgrid velocity field is used to represent only small scales by definition, and hence, its  covariance function  corresponds to the interaction of only  small scales. This is matched with the literal definition of  Reynolds stress. The covariance at space and time lag (0,0) is used as an approximation for the Reynolds stress $R_{ij}$  by
\begin{equation}\nonumber
{{R}^{ij}}(0,0)=\mathbb{E}[{{{u}'}_{i}}(r,s){{{u}'}_{j}}(r,s)]\sim
\overline{{{{{u}'}}_{i}}{u}'_{j}}={{R}_{ij}}
\end{equation}
 For \c{C}inlar velocity field, we get
 \begin{align}\label{15}
R_{ij}\equiv {{R}^{ij}}(0,0)&=\frac{\lambda }{2c}\int_{\mathbb{R}}{\alpha
}(da){{a}^{2}}\int_{{{\mathbb{R}}^{2}}}{d}z\int_{{{\mathbb{R}}^{+}}}{\beta
}(db)\frac{{{b}^{2}}}{|z{{|}^{2\gamma }}}{{\upsilon
}_{i}}(z){{\upsilon }_{j}}(z) \nonumber\\
&=\frac{\lambda
}{2c}\mathbb{E}({{a}^{2}})\mathbb{E}({{b}^{2}})\int_{{{\mathbb{R}}^{2}}}dz\,{\frac{{{\upsilon
}_{i}}(z){{\upsilon }_{j}}(z)}{|z{{|}^{2\gamma }}}}
\end{align}
	Substituting the basic vortex \eqref{basicvelocity}  in \eqref{15}, we obtain the Reynolds stress  as
\begin{equation}\label{ReynoldsStress}
{{R}_{ij}}\equiv {{\delta }_{ij}}\frac{\lambda
}{c}\mathbb{E}({{a}^{2}})\mathbb{E}({{b}^{2}})\frac{3\pi }{16}
\end{equation}
where we have taken $\gamma =1/2$ for simplifying the result. The  Reynolds stress $R_{ij}$ will be  parameterized as described below.

\subsection{Modelling the Parameters of Reynolds Stress}

Our aim is to represent Reynolds stress, which captures the fluctuations of the subgrid scale velocity, in terms of the resolved velocity field.  The generation of small-scale fluctuations is due to the nonlinear term in the equation of motion.  However, the viscous terms prevent the generation of infinitely small scales of motion by dissipating small-scale energy into heat and smoothing out the velocity fluctuations \cite{tennekes}. For flows with high Reynolds number, the turbulent kinetic energy, that is generated  at large scales,  cascades to smaller scales and then dissipates in the viscous range. On the other hand, the viscous stress depends linearly on the strain rate \cite{pope}. Therefore, the dissipation rate is directly proportional to the strain rate, which is expected to increase with the wave number.

The strain rate causes the deformation of eddies shape and local dissipation \cite{frisch}. The fluid elements are extended or contracted in the straining motion. We describe  a representation for each parameter appearing in the Reynolds stress \eqref{ReynoldsStress} using these properties of the strain rate. Depending on the meaning of a parameter,
we refer to viscosity, the dissipation rate, and the strain rate, interchangeably, as they are proportional to each other. Our analysis is clearly inspired by eddy viscosity models, in which the deviatoric part of the subgrid scale stress is modelled as a linear function of the strain rate tensor. However, \eqref{ReynoldsStress} and representation of its parameters involve the aspects of vortex formation and decay as well.

\subsubsection*{Decay rate $c$}
 The eddy viscosity causes the energy dissipation. In the original \c{C}inlar velocity field model $\tilde{u}$ \cite{cinlar1993},  small
eddies are dissipated by decay rate $c$, due to the following equation
$$
d\tilde{u}(x,t)=-c\; \tilde{u}(x,t)dt+\int_Q N(dt,dq).
$$
that  $\tilde{u}$ satisfies. This equation does not hold with the generalized form $c_q$, but we use the above equation to capture the essence of the decay rate.
Therefore, the parameter $c$  is approached as eddy viscosity, or the dissipation rate. The  eddy viscosity $\nu_t$ modelled by  Smagorinsky \cite{sma63} is proportional to the filter width and the magnitude of the strain rate. Similarly, the dissipation rate is directly proportional to the strain rate. Therefore, we set
 $$c \equiv C_1 |\bar{S}|$$
where
\begin{equation}\nonumber
|\bar{S}|=2(\bar{S}_{ij}\bar{S}_{ij})^{1/2}
\end{equation}
and $C_1>0$ is a constant.

\subsubsection*{Shape parameter $\theta$ and scale parameter $\zeta$ }
In incompressible flows, the  strain rate affects the shape of eddies and  leads to  their splitting  into two or more smaller ones \cite[pg.260]{tennekes}. Therefore, we also take the radius
$b$ as inversely proportional to $|\bar{S}|$. The expected value of radius $b$ is calculated using right truncated Gamma distribution. We get
\begin{equation}\nonumber
            \mathbb{E}({{b}^{2}})=-\frac{{{B}^{\theta }}{{e}^{-B/\zeta
}}}{{{\zeta }^{\theta -2}}{{\Gamma }_{B/\zeta }}(\theta
)}(B/\zeta +\theta +1)+\theta (\theta +1){{\zeta }^{2}}, \quad \quad 0 < b <B
			\end{equation}
			
The shape parameter $\theta$ is unit-less and the unit of the scale parameter is characteristic length scale $L$. While $|\bar{S}|$ increases, smaller eddies emerge.  Strain rate $|\bar{S}|$ affects directly the shape parameter $\theta$, so we can model $\theta\propto 1/|\bar{S}|T$, where $T$ indicates the characteristic time and it can be taken as $T=\bar{\Delta}/\bar{u}$. That is, $\theta$ is modelled by  $$\theta \equiv C_2\frac{\bar{\Delta}}{|\bar{S}|\bar{u}}$$   where $C_2 > 0$ is a model constant. Also the change of scale parameter $\zeta$ only affects the range of the radius distribution, which is proportional to $\zeta$. This linear relationship between $\zeta$ and $B$ can be written as $\zeta \equiv C_3B $, where $C_3>0$ is a constant.

\subsubsection*{Arrival rate $\lambda$}
The arrival rate $\lambda$ is defined as number of eddies per unit area and time. Therefore, its dimension is ${1}/{\left(T{{L}^{2}} \right)}\;$.
Due to occurrence of new small eddies as a result of the strain rate, the number of eddies per unit area and time in subgrid scale  increases. This implies that  $ \lambda $ is proportional to the strain rate. Using this information and dimension analysis, we get
 $$\lambda \equiv C_4\frac{|{{{\bar{S}}}}|}{{{\bar{\Delta} }^{2}}}$$ where $C_4>0$ is a model constant.

\subsubsection*{Expectation of $a^2$}
Lundgren and Burgers \cite{burgers, lundgren1982} assume that the radial velocity decreases linearly with the  strain rate. However, in \c{C}inlar velocity field \eqref{velocity}, the radial velocity magnitude, which is described by $a$, decreases exponentially in
time. We note this by the term
\[
e^{-c_q(t-s)}\,a\, v_q(x).
\]
 Then, the radial velocity simply becomes $e^{-c}a$ after a unit time increment $t-s\equiv 1$, from the initial magnitude $a$ of the arriving vortex $v_q$. Clearly, the square of the initial magnitude $a$ decays with the rate $e^{-2c}$.

 At each time step of LES, we assume that the initial velocity at the beginning of this time step, namely $\bar{u}$ acts as a proxy to an average value for the magnitude $a$ of each arriving vortex in the subgrid scale. Then, since the magnitude would decay with the rate $e^{-c}$ as explained above, we can model its square $\mathbb{E}(a^2)$ as $\bar{u}^2 e^{-2c}$ for a unit time of decay.
 Therefore, the second moment of $a$ is taken to be proportional to the  exponential of the strain rate as
 $$E(a^2)\equiv \bar{u}^2 e^{-2C_1|\bar{S}|}$$ 
in view of the approximation $c \equiv C_1|\bar{S}|$.

\vspace{5mm}

Using  our arguments above, we get the model for the Reynolds stress as
\begin{equation}\label{ReynoldsModel}
{{R}_{ij}}\equiv \delta_{ij}\frac{3\pi}{64}\frac{C_3^2C_4}{C_1}\bar{u}^2\bar{\Delta}^2e^{-2C_1|\bar{S}|}
\left[\theta\left( \theta+1\right) - \frac{C_3^{-\theta}}{e^{1/C_3}\Gamma_{1/C_3}\left(\theta\right)}\left(\frac{1}{C_3}+\theta+1\right) \right]
\end{equation}
where the radius  of the largest eddy in dissipation range $ B$ is taken to be  equal to half of the grid size as $\bar{\Delta}/2 $, and $C_2\bar{\Delta}/(|\bar{S}|\bar{u})$ will be used for $\theta$ as discussed above.

\section{Numerical Results and Comparison}

In this section, the channel flow results of the LES simulation with  three different SGS models, namely, \c{C}inlar, Smagorinsky and one equation eddy, are compared with the DNS performed by Moser et al. \cite{moser} for friction Reynolds numbers of 395 and 590, and by Hoyas and Jimenez \cite{jimenez} for a friction Reynolds number of 950. The friction Reynolds number is defined as $Re_\tau = u_\tau \delta/\nu$ where $u_\tau =\tau_\omega/\rho$ is the friction velocity, $\tau_\omega$ is  the wall shear stress and  $\delta$ is the channel half height. Fully developed  channel flow has been studied extensively to increase the understanding of the mechanics of wall-bounded turbulent flows and it is a baseline for validation of a turbulence model. The periodic boundary condition in the streamwise and spanwise directions, and no-slip boundary condition on the wall have been applied.

LES is performed by using the OpenFOAM CFD Toolbox \cite{openfoam}. A finite-volume based method is used for  numerical calculations in OpenFOAM LES solver. The PIMPLE algorithm is used for the pressure-velocity coupling. For the pressure, the Poisson equation  is solved using an algebraic multi-grid (AMG) solver. When the scaled residual becomes less than $10^{-6}$ , the algebraic equation is considered to have converged. Using adjustable time step, the time step has been modified dynamically to guarantee a constant Courant  number of $0.2$. The computational mesh are $128\times  98\times 128$ for $Re_{\tau} = 395, 590$ and  $128\times  128\times 128$  for $Re_{\tau} = 950$. The box size is $2\pi\delta \times 2\delta \times \pi\delta$ for the
streamwise, wall-normal and spanwise directions, respectively.

The numerical results are depicted through graphs of the  time and space averaged quantities normalized by the friction velocity $u_\tau$:
\begin{itemize}
\item the mean streamwise velocity $\langle \bar{u}/u_\tau \rangle$,
\item the $x$, $y$ component of the Reynolds stress $\langle u'v' \rangle/u_\tau^2$,
\item the  mean squared (ms) velocity fluctuations given by the streamwise $\langle u'u'\rangle /u_\tau^2$, wall-normal $\langle v'v'\rangle /u_\tau^2$, and spanwise $\langle w'w' \rangle/u_\tau^2$ quantities,
\end{itemize}
where $\langle \cdot \rangle$ denotes time and space averaging, the fluctuating quantities $f'$  are calculated as $f'=f-\langle f \rangle$. In the graphs, a $"+"$ sign denotes that the variable is normalized with $u_\tau$, as above. For example,  $uu+$ corresponds to $\langle u'u'\rangle /u_\tau^2$.
For compatibility with Jimenez data at $Re_{\tau} = 950$, square root is taken for the streamwise, wall-normal and spanwise velocity fluctuations before comparison.

\subsection{Results for $Re_\tau =395$}

\begin{figure}
    \centering
    \begin{subfigure}[b]{0.45\textwidth}
        \includegraphics[width=\textwidth]{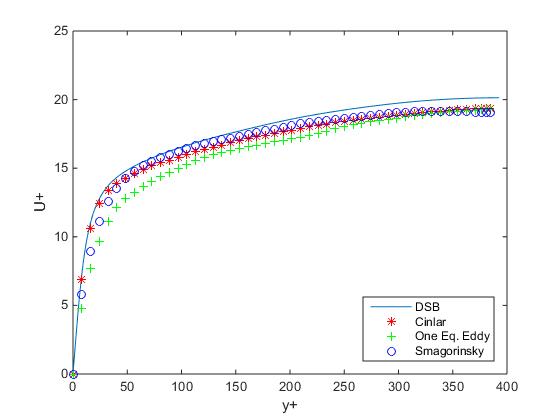}
        \caption{}
        \label{fig:velocityRe395Tez}
    \end{subfigure}
    ~ 
    \begin{subfigure}[b]{0.45\textwidth}
        \includegraphics[width=\textwidth]{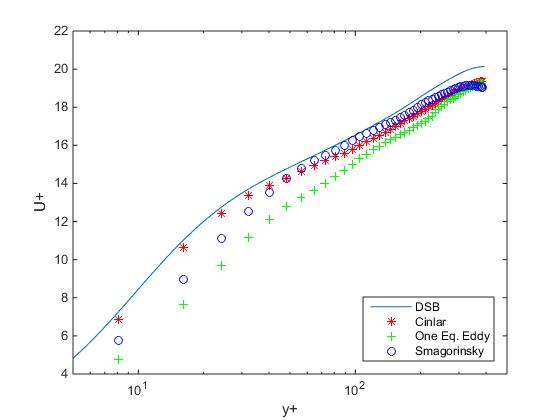}
        \caption{}
        \label{fig:Re395logxU+}
    \end{subfigure}
    ~ 
    \caption{Channel flow comparison data for $Re_\tau=395$, (a) mean velocity profile, (b) mean velocity profile for logarithmic scale }\label{fig:Re395Mean}
\end{figure}

In the first test case, we compare LES results with Moser DNS data for $Re_\tau=395$ \cite{moser}. The mean streamwise velocity is given in Fig. \ref{fig:Re395Mean}, where the superscript $+$ denotes non-dimensionalized quantities with the friction velocity $u_\tau$. In particular, we have
\[
U+=\langle U_1 \rangle/u_\tau \; , \quad \quad \quad y+ = yu_\tau /\nu \; .
\]
The mean velocity profile with \c{C}inlar SGS model is in good agreement with  DNS results of Moser et.al. and LES computations of Smagorinsky  and one equation eddy models. \c{C}inlar and  Smagorinsky models show the best fit regarding the mean streamwise velocity. Especially, \c{C}inlar SGS model provides better results  in the viscous wall region  $(0<y+<50)$.

\begin{figure}
    \centering
    \begin{subfigure}[b]{0.45\textwidth}
        \includegraphics[width=\textwidth]{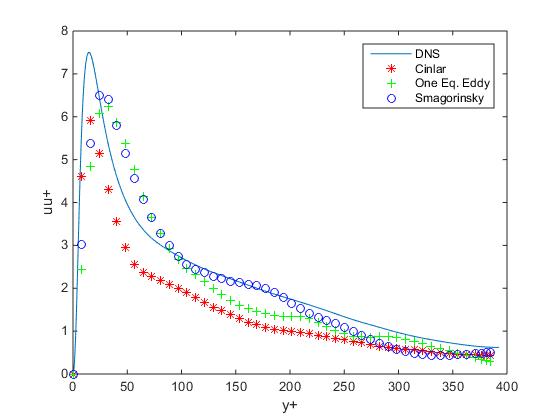}
        \caption{}
        \label{fig:uuGraphRe395Tez}
    \end{subfigure}
    ~ 
    \begin{subfigure}[b]{0.45\textwidth}
        \includegraphics[width=\textwidth]{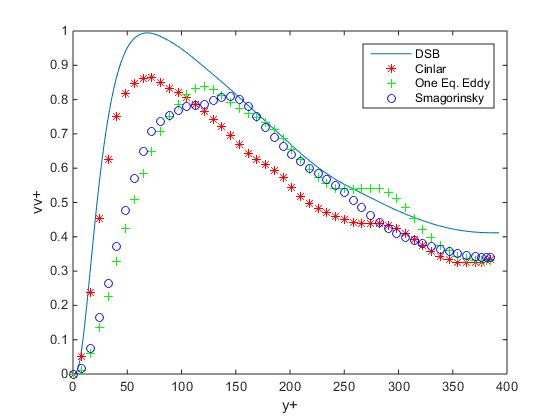}
       \caption{}
        \label{fig:vvGraphRe395Tez}
    \end{subfigure}
    \newline
    \begin{subfigure}[b]{0.45\textwidth}
        \includegraphics[width=\textwidth]{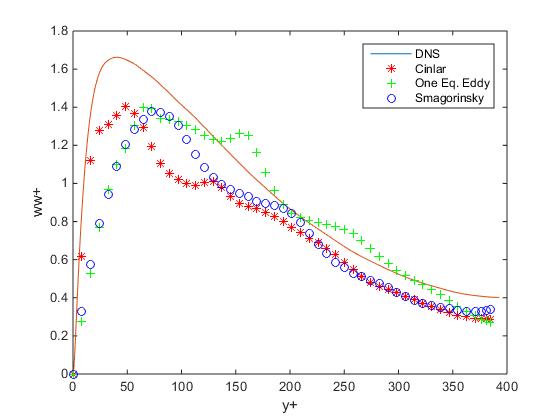}
        \caption{}
        \label{fig:wwGraphRe395Tez}
    \end{subfigure}
    ~ 
    \begin{subfigure}[b]{0.45\textwidth}
        \includegraphics[width=\textwidth]{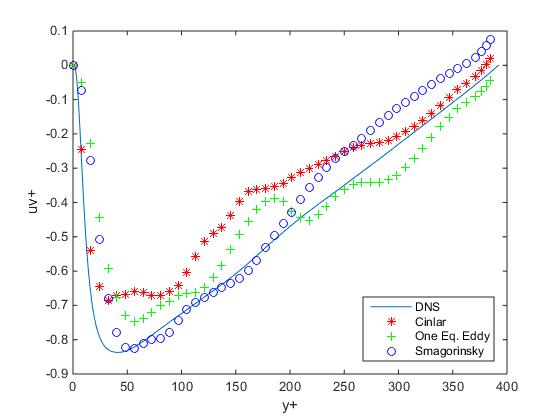}
        \caption{}
        \label{fig:uvGraphRe395Tez}
    \end{subfigure}
    \caption{Channel flow comparison data for $Re_\tau=395$,  (a) ms velocity fluctuation streamwise direction; (b) ms velocity fluctuation, wall-normal direction, (c) ms velocity fluctuation  spanwise direction, (d) shear stress $u'v'+$ velocity profiles}\label{fig:Re395Stress}
\end{figure}

In  Fig. \ref{fig:Re395Stress}, the mean square (ms)  velocity fluctuations are plotted. All models predict the velocity fluctuations quite accurately. Especially in the viscous subregion, which has  poor resolution for  LES  compared with DNS,  LES results with \c{C}inlar SGS model are remarkably good. Our model leads to over or under-prediction of the  ms values from its peak to the outer layer of the channel flow where the viscosity is not prevalent. The other models also deviate from DNS, but in different regions. In particular, the value of $y+$ where ms velocities reach their peak values is best predicted by our model. Our results agree with those obtained with Smagorinsky model towards outer region, except for  Fig. \ref{fig:Re395Stress} d), where \c{C}inlar SGS model performs better.

\subsection{Results for $Re_\tau =590$}

\begin{figure}
    \centering
    \begin{subfigure}[b]{0.45\textwidth}
        \includegraphics[width=\textwidth]{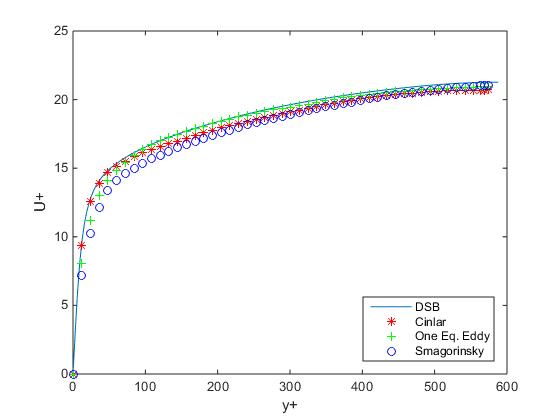}
        \caption{}
        \label{fig:velocityRe590Tez}
    \end{subfigure}
    ~ 
    \begin{subfigure}[b]{0.45\textwidth}
        \includegraphics[width=\textwidth]{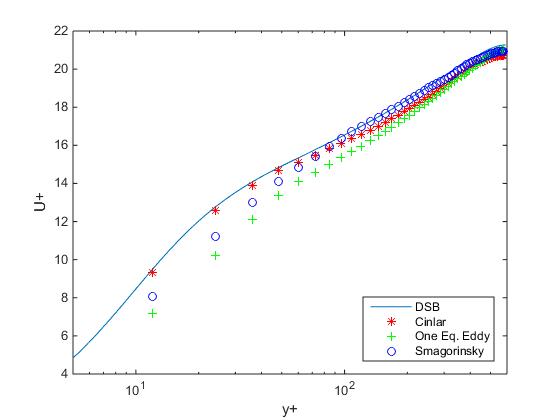}
        \caption{}
        \label{fig:Re590logxU+}
    \end{subfigure}
    ~ 
    \caption{Channel flow comparison data for $Re_\tau=590$, (a) mean velocity profile, (b) mean velocity profile for logarithmic scale }\label{fig:Re590Mean}
\end{figure}

For c, mean streamwise velocity profiles are shown for DNS data and LES with the three SGS models in Fig. \ref{fig:Re590Mean}. It can be seen from the mean velocity profile graphs, \c{C}inlar model yields the most accurate result in the viscous region. Smagorinsky model gives better approximation than one equation eddy overall, and better results towards the outer layer of the channel flow.

\begin{figure}
    \centering
      \begin{subfigure}[b]{0.45\textwidth}
        \includegraphics[width=\textwidth]{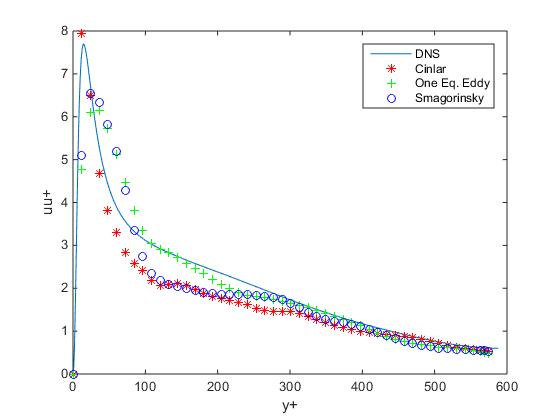}
        \caption{}
        \label{fig:uuGraphRe590Tez}
    \end{subfigure}
    ~ 
    \begin{subfigure}[b]{0.45\textwidth}
        \includegraphics[width=\textwidth]{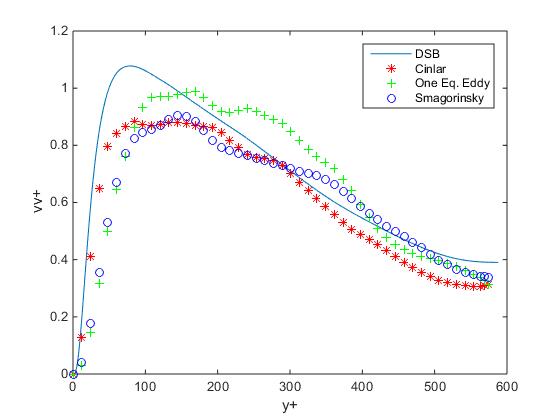}
        \caption{}
        \label{fig:vvGraphRe590Tez}
    \end{subfigure}
    \newline
    \begin{subfigure}[b]{0.45\textwidth}
        \includegraphics[width=\textwidth]{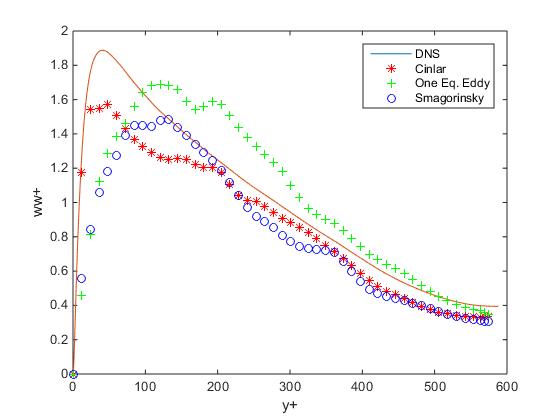}
        \caption{}
        \label{fig:wwGraphRe590Tez}
    \end{subfigure}
    ~ 
    \begin{subfigure}[b]{0.45\textwidth}
        \includegraphics[width=\textwidth]{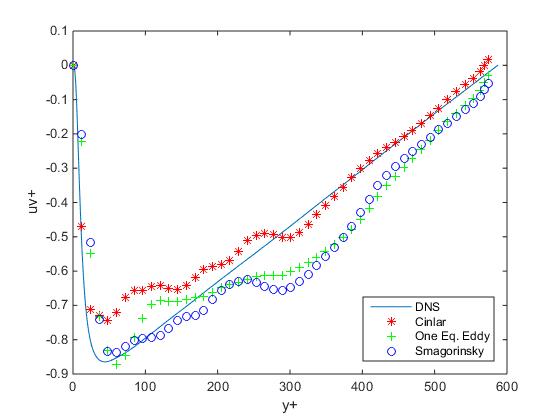}
        \caption{}
        \label{fig:uvGraphRe590Tez}
    \end{subfigure}
    \caption{Channel flow comparison data for $Re_\tau=590$,  (a) ms velocity fluctuation  streamwise direction; (b) ms velocity fluctuation wall-normal direction, (c) ms velocity fluctuation  spanwise direction, (d) shear stress $u'v'+$ velocity profiles}\label{fig:Re590Stress}
\end{figure}

Fig. \ref{fig:Re590Stress} shows comparison of ms velocity fluctuations with DNS data and LES results. LES results with \c{C}inlar model match DNS data better than the other models in the viscous range and coincides with Smagorinsky model towards outer layer. The fluctuations in the three directions shown in Fig. \ref{fig:Re590Stress} a)-c) attain  slightly lower values than DNS data, and the shear stress is lower in magnitude as well for $y+$ between 50 and 300. However, the shear stress is best approximated for $y+>300$ by our model, like the viscous range.  There is little discrepancy with  the peak values of DNS for our model whereas the other models produce graphs which look somewhat shifted to the  right.

 \subsection{Results for $Re_\tau=950$}

\begin{figure}
    \centering
    \begin{subfigure}[b]{0.45\textwidth}
        \includegraphics[width=\textwidth]{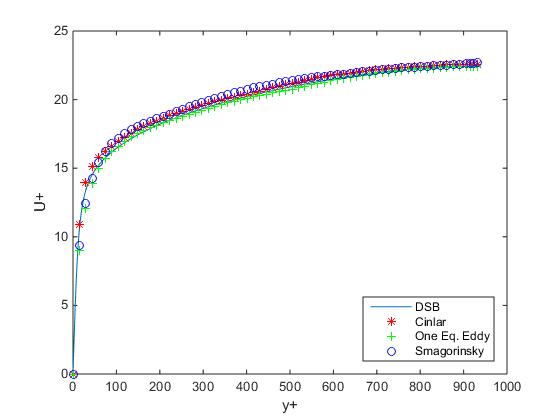}
        \caption{}
        \label{fig:velocityRe950Tez}
    \end{subfigure}
    ~ 
    \begin{subfigure}[b]{0.45\textwidth}
        \includegraphics[width=\textwidth]{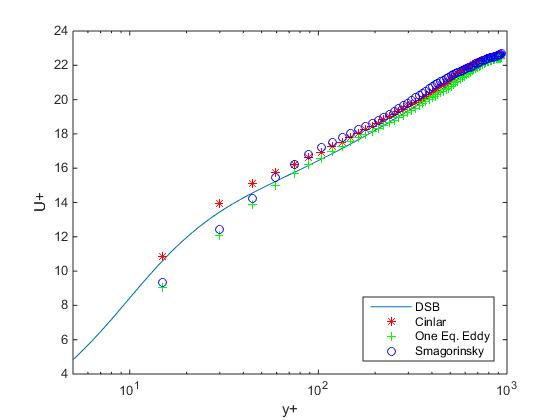}
        \caption{}
        \label{fig:Re950logxU+}
    \end{subfigure}
    ~ 
    \caption{Channel flow comparison data for $Re_\tau=950$, (a) mean velocity profile, (b) mean velocity profile for logarithmic scale }\label{fig:Re950Mean}
\end{figure}

The last comparison is for $Re\tau=950$ for validating \c{C}inlar model. The velocity profiles are shown in Fig. \ref{fig:Re950Mean}. It can been seen that \c{C}inlar model over-predicts the mean velocity profile throughout the channel, but with clearly less error than Smagorinsky and one equation eddy models.

\begin{figure}
    \centering
       \begin{subfigure}[b]{0.45\textwidth}
        \includegraphics[width=\textwidth]{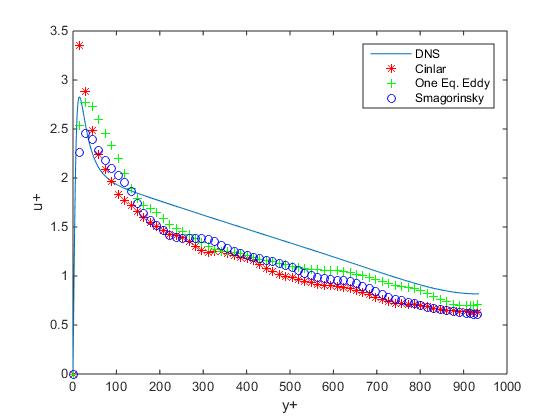}
        \caption{}
        \label{fig:uuGraphRe950Tez}
    \end{subfigure}
    ~ 
    \begin{subfigure}[b]{0.45\textwidth}
        \includegraphics[width=\textwidth]{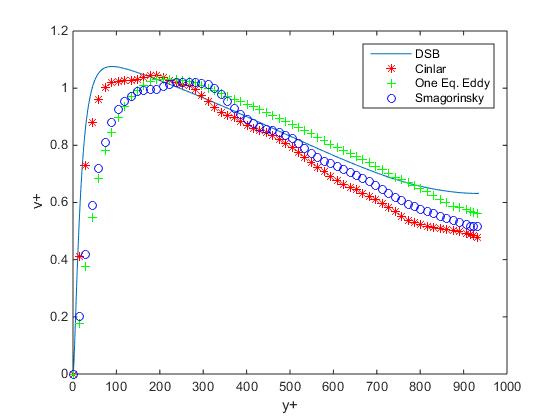}
        \caption{}
        \label{fig:vvGraphRe950Tez}
    \end{subfigure}
    \newline
    \begin{subfigure}[b]{0.45\textwidth}
        \includegraphics[width=\textwidth]{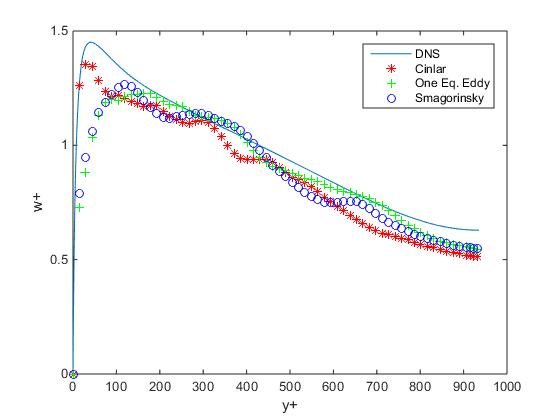}
        \caption{}
        \label{fig:wwGraphRe950Tez}
    \end{subfigure}
    ~ 
    \begin{subfigure}[b]{0.45\textwidth}
        \includegraphics[width=\textwidth]{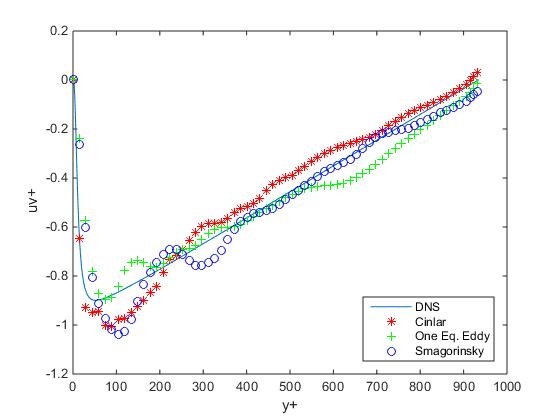}
        \caption{}
        \label{fig:uvGraphRe950Tez}
    \end{subfigure}
    \caption{Channel flow comparison data for $Re_\tau=950$,  (a) rms velocity fluctuation  streamwise direction; (b) rms velocity fluctuation wall-normal direction, (c) rms velocity fluctuation spanwise direction, (d) shear stress $u'v'+$ velocity profiles
}\label{fig:Re950Stress}
\end{figure}

Fig. \ref{fig:Re950Stress} presents the rms velocity fluctuations as the second-order turbulent statistics   against the rms profiles from a DNS simulation of channel flow at $Re_\tau=950$ \cite{jimenez}.  The results for the three different SGS are not significantly different from each other. All SGS models capture the general rms profile of the DNS data while \c{C}inlar model still behaves better in the viscous range and in predicting the position of the peak values in some cases. Hence, we see that it is valid also for the high Reynolds number case.

\section{Conclusion}

In this paper, we have modelled the Reynolds stress tensor of the generalized \c{C}inlar random velocity field, which was shown to represent Eulerian velocity of the subscales accurately in previous work. Because an analytical expression is available for  Reynolds stress, we have represented its parameters originating from  the probability distributions with the resolved velocity field, in particular as functions of the resolved strain rate tensor.

Our numerical results demonstrate that LES of
fully  developed  turbulent  channel  flow  with \c{C}inlar SGS model is  in remarkably good
agreement  with the available DNS data for Reynolds numbers $395, 590$ and $950$, by comparison with benchmark models, namely Smogorinsky and one equation eddy.
\c{C}inlar model yields especially better results in the viscous subregion near the wall, which has  poor resolution for  LES  compared with DNS. The computational burden is much less than one equation eddy and is observed to be as low as Smagorinsky in simulations.

As future work, \c{C}inlar velocity field can be extended to $\mathds{R}^3$ where the basic eddy can be chosen to be the unit sphere, in analogy with the unit
disk used in two dimensions, and the planar motion can be taken as a rotation.

\vspace{7mm}

\noindent{\bf Acknowledgements}. This work was supported by The Scientific and Technological Research Council of Turkey (TUBITAK) Project No. 112T761.  The numerical calculations reported in this paper were partially performed at TUBITAK ULAKBIM, High Performance and Grid Computing Center (TRUBA resources). The authors would like to thank Ay\c{s}e G\"{u}l  G\"{u}ng\"{o}r, Alkan Kabak\c{c}\i o\u{g}lu,  and Hasret T\"{u}rkeri for their helpful discussions on the physics and simulations of the flow.



\end{document}